 \documentclass[final,5p,times,twocolumn,number,sort&compress]{elsarticle}

\usepackage{amssymb}
\usepackage{lipsum}
\usepackage{amsthm}

\usepackage{dcolumn}
\usepackage{bm}
\usepackage[hidelinks,colorlinks=true,linkcolor=blue, citecolor=blue, urlcolor=blue]{hyperref}
\hypersetup{
  pdftitle={SO3NF},
  pdfauthor={Tokuro Fukui}
}
\usepackage{url}
\usepackage{fancybox}
\usepackage{geometry}
\usepackage{wrapfig}
\usepackage{amsmath,cases}
\usepackage{mathtools}
\usepackage{amssymb}
\usepackage{bbold}
\usepackage{tocloft}
\usepackage{mathrsfs}
\usepackage{braket}
\usepackage{longtable}
\usepackage{dcolumn}
\usepackage{multirow}
\usepackage{framed}
\usepackage{arydshln}
\usepackage{chngcntr}
\usepackage{color}
\usepackage{booktabs}
\usepackage{tabularx}
\usepackage{array}
\usepackage{ulem}
\usepackage{cancel}

\newcolumntype{Y}{>{\centering\arraybackslash}X} 
\newcolumntype{Z}{>{\raggedright\arraybackslash}X} 
\newcolumntype{L}{>{\raggedright\arraybackslash}p{1cm}} 
\newcolumntype{W}{>{\centering\arraybackslash}p{2cm}}   
\newcommand{\vect}{\boldsymbol}

\journal{Physics Letters B}

\begin{document}

\begin{frontmatter}



\title{Uncovering the mechanism of chiral three-nucleon force in driving spin-orbit splitting}


\author[aff1,aff2]{Tokuro Fukui}
\ead{tokuro.fukui@artsci.kyushu-u.ac.jp}
\address[aff1]{Faculty of Arts and Science, Kyushu University, Fukuoka 819-0395, Japan}

\address[aff2]{RIKEN Nishina Center, Wako 351-0198, Japan}

\author[aff3,aff4]{Giovanni De Gregorio}
\ead{degregorio@na.infn.it}
\address[aff3]{Dipartimento di Matematica e Fisica, Universit\`{a} degli Studi della Campania ``Luigi Vanvitelli'', viale Abramo Lincoln, Caserta 5-I-81100, Italy}

\address[aff4]{Istituto Nazionale di Fisica Nucleare, Complesso Universitario di Monte S. Angelo, Via Cintia, Napoli I-80126, Italy}

\author[aff4]{Angela Gargano}
\ead{agargano@na.infn.it}

\begin{abstract}
The three-nucleon force (3NF) is crucial in shaping the shell structure of atomic nuclei, 
particularly impacting the enhancement of spin-orbit (SO) splitting, 
especially in nuclei with significant deviations from stability. 
Despite its importance, the specific mechanisms driving this enhancement remain unclear. 
In this study, we introduce a decomposition scheme based on the rank of irreducible tensors forming the 3NF, 
derived from chiral effective field theory at next-to-next-to-leading order, 
to elucidate their influence on SO splitting. 
Within the shell-model framework, our analysis reveals that the rank-1 component of the 3NF is the primary factor 
enlarging the energy gap between the $0p_{3/2}$ and $0p_{1/2}$ single-particle levels in $p$-shell nuclei. 
Since the rank-1 component originates exclusively from the $2\pi$-exchange 3NF,
our finding will not depend on the choice of the low-energy constants of contact terms.
We also remark on the antisymmetry of the rank-1 3NF, 
which can affect the quantum entanglement of spin states. 
This study lays the groundwork for further exploration into this field 
toward a microscopic understanding of the 3NF impact on the nuclear shell structure.
\end{abstract}



\begin{keyword}
Three-nucleon force \sep Spin-orbit splitting \sep Chiral effective field theory \sep Shell model



\end{keyword}

\end{frontmatter}




\section{Introduction}
\label{secIntro}
The understanding of the structure and dynamics of atomic nuclei hinges not only on the intricate nature of two-nucleon forces (2NFs) 
but also on the significant impact of three-nucleon forces (3NFs) 
(see for instance Ref.~\cite{annurev:/content/journals/10.1146/annurev-nucl-102313-025446}). 
Since Primakoff and Holstein~\cite{PhysRev.55.1218} argued for the importance of many-body forces in quantum systems, 
considerable efforts have been dedicated to studying 3NFs in nuclear physics. 
Initiating with the Fujita--Miyazawa force~\cite{10.1143/PTP.17.360}, 
various methods have been proposed to construct 3NFs based on pion exchanges 
(see Refs.~\cite{doi:10.1146/annurev.ns.34.120184.002155,PhysRevC.59.53} for the comparison of 3NFs). 
Recent advancements rooted in studies by Weinberg~\cite{WEINBERG1992114} 
and van Kolck with coauthors~\cite{PhysRevC.49.2932,PhysRevLett.85.2905} on 3NFs from a chiral Lagrangian 
have notably progressed with the advent of chiral effective field theory (EFT)~\cite{Weinberg1979327,Epelbaum2006654,MACHLEIDT20111}.
This theory offers a systematic and unified approach to constructing 2NFs and 3NFs on equal footing~\cite{WEINBERG1992114,PhysRevC.49.2932,MACHLEIDT20111}. 

These modern 2NFs and 3NFs have been extensively utilized in many-body calculations, 
firmly establishing the role of 3NFs in describing 
spectroscopic properties~\cite{PhysRevLett.99.042501,PhysRevC.87.014327,PhysRevLett.113.262504}, 
scattering processes~\cite{PhysRevC.88.054622,PhysRevC.91.021301,PhysRevC.102.024616} of light systems,
neutron drip-line locations of oxygen isotopes~\cite{PhysRevLett.105.032501,PhysRevLett.108.242501,PhysRevLett.111.062501,MA2020135257,ZHANG2022136958}, 
and fundamental properties of medium-heavy nuclei~\cite{PhysRevLett.110.022502,PhysRevC.90.024312,PhysRevC.100.034324,PhysRevLett.130.192501}.
As concerns the latter two points, 
we refer the readers to a latest review on the chiral 3NF studied within the shell model~\cite{CORAGGIO2024104079}. 
Moreover, 3NFs essentially contributes to
saturation properties of nuclear matter~\cite{PhysRevC.88.064005,PhysRevC.89.044321,doi:10.1142/S0218301317300016,10.3389/fphy.2019.00213,PhysRevC.102.034313,PhysRevC.104.064312}.

Notably, attention has been drawn to the relationship between chiral 3NFs 
and variations in shell structure observed when transitioning from stable to exotic nuclei. 
These 3NFs are found to be crucial in explaining the nonuniversality of magic numbers~\cite{PhysRevLett.105.032501,Holt_2012,Holt_2013,HoltEPJA,PhysRevC.95.021304,RevModPhys.92.015002,MA2020135257}, 
highlighting their potential to influence the spin-orbit (SO) splitting, 
a key aspect of atomic nuclei's shell structure manifestation. 
The contribution of 3NFs to SO splitting has been investigated even before the emergence of chiral EFT, 
with historical discussions providing insights into their microscopic origins and experimental prospects 
(see Refs.~\cite{10.1143/PTP.17.366,10.1143/PTP.66.227,PhysRevLett.70.2541} for prior discussions on the relation between SO splitting and 3NFs, 
as well as Ref.~\cite{Uesaka2016} for a comprehensive historical overview, including experimental study prospects).
\begin{figure}[!t]
\begin{center}
 \includegraphics[width=0.5\textwidth,clip]{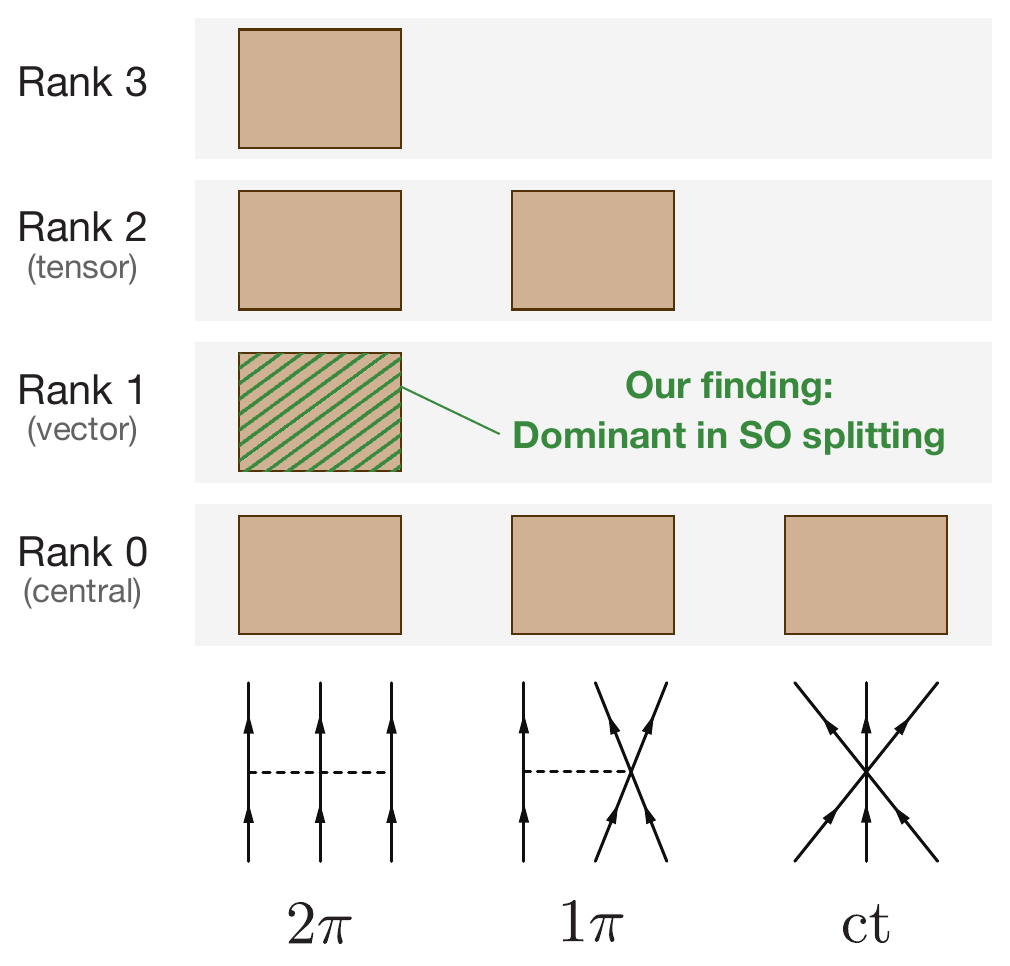}
 \caption{Classification of the chiral 3NF at N$^2$LO by the number of exchanged pions and rank of the irreducible tensors.
 The main finding of this Letter is that the rank-1 3NF originating from the $2\pi$-exchange process 
 crucially contributes to the SO splitting of light nuclei. See texts for details.}
\label{fig3NFrank}
\end{center}
\end{figure}

The relevance of the chiral 3NF in the enhancement of the SO splitting was discussed within
the shell model~\cite{PhysRevLett.105.032501,Holt_2012,PhysRevC.95.021304,PhysRevC.98.044305,PhysRevC.100.034324,MA2020135257},
density functional theory using the density-matrix expansion~\cite{PhysRevC.68.054001},
and nuclear matter calculations relying on the Brueckner theory~\cite{PhysRevC.86.061301}.

However, the specific mechanism through which the chiral 3NF influences the SO splitting remains elusive.
A crucial insight to address this issue lies in characterizing the 3NF in terms of its tensorial components 
and evaluating the relative weight of these components in determining the SO splitting.

A pioneering work in studying the 3NF components within the tensor space was performed by And\={o} and Band\={o} in 1981~\cite{10.1143/PTP.66.227}.
They reported that the SO splitting in $^{16}\mathrm{O}$ and $^{40}\mathrm{Ca}$ is enhanced by the rank-1 tensor of 3NFs.
However, the influence of the other tensors was not explored 
since they solely adopted the rank-1 component of
the Fujita--Miyazawa~\cite{10.1143/PTP.17.360} and Tucson--Melbourne forces~\cite{COON1979242}.

In this Letter, we propose a decomposition scheme of the chiral 3NF at next-to-next-to-leading order (N$^2$LO)~\cite{PhysRevC.66.064001} 
by the rank of the irreducible tensors,
leading to a categorization of the 3NF in terms of the number of exchanged pions and rank, as illustrated in Fig.~\ref{fig3NFrank}.
Subsequently, as a test case to investigate the role of the different rank components,
we select $p$-shell nuclei and demonstrate that the enhancement of the SO splitting resulting from the 3NF is mainly related to its rank-1 tensor.

\section{Method}
\label{secMethod}
The chiral 3NF at N$^2$LO contains three terms;
the two-pion ($2\pi$) exchange term, one-pion ($1\pi$) exchange plus contact term, and contact (ct) term,
expressed by the left, middle, and right diagrams of Fig.~\ref{fig3NFrank}, respectively.
Furthermore, the $2\pi$ term comprises three distinct constituents 
associated with the pion-nucleon low-energy constants (LECs), $c_1$, $c_3$, and $c_4$,
which also appear at the 2NF level.
The potential of each term of the 3NF can be schematically written as
\begin{align}
 v_{3N}^{(\alpha)}
 &= \sum_{i \ne j \ne k} v^{(\alpha)}\!\left(\vect{\tau}_i,\vect{\tau}_j,\vect{\tau}_k\right)
 w^{(\alpha)}\!\left(\vect{\sigma}_i,\vect{\sigma}_j,\vect{\sigma}_k,\vect{q}_i,\vect{q}_j\right),
 \label{3Npotgeneral}
\end{align}
where $v^{(\alpha)}$ is the isospin part and $w^{(\alpha)}$ represents the spin-momentum dependent part.
The letter $\alpha$ stands for ct, $1\pi$, or one of the three constituents of the $2\pi$ term,
and the indices $i, j, k  \in \{1,2,3\}$ specify the nucleon.
The spin and isospin operators of the nucleon ${i}$ are denoted by 
the Pauli matrices, $\vect{\sigma}_i$ and $\vect{\tau}_i$, respectively.
The momentum transfer is given by ${\vect{q}_i=\vect{p}_i'-\vect{p}}$, where $\vect{p}_i$ ($\vect{p}_i'$) is
the initial (final) momentum of the nucleon $i$.

Herein, we depict the general structure of the irreducible-tensor decomposition of $w^{(\alpha)}$,
which can be written as
\begin{align}
 &w^{(\alpha)}\!\left(\vect{\sigma}_i,\vect{\sigma}_j,\vect{\sigma}_k,\vect{q}_i,\vect{q}_j\right)
 \nonumber\\
 &\quad=
 w_{\mathrm{pro}}^{(\alpha)}(q_i,q_j)
 \sum_{\lambda}
 \mathcal{O}_\lambda^{(\alpha)}\!\left(\vect{\sigma}_i,\vect{\sigma}_j,\vect{\sigma}_k,\hat{\vect{q}}_i,\hat{\vect{q}}_j\right),
 \label{3Npotsigq}\\
 &\mathcal{O}_\lambda^{(\alpha)}\!\left(\vect{\sigma}_i,\vect{\sigma}_j,\vect{\sigma}_k,\hat{\vect{q}}_i,\hat{\vect{q}}_j\right)
 \nonumber\\
 &\quad=
 A_\lambda
 \left[\mathcal{M}_{\lambda}^{(\alpha)}\!\left(\vect{\sigma}_i,\vect{\sigma}_j,\vect{\sigma}_k\right)
 \otimes
 \mathcal{N}_{\lambda}^{(\alpha)}\!\left(\hat{\vect{q}}_i,\hat{\vect{q}}_j\right)\right]_{00}.
 \label{3NpotOlam}
\end{align}
The irreducible tensors of rank $\lambda$ of the spin and momentum parts are expressed by 
$\mathcal{M}_{\lambda}^{(\alpha)}$ and $\mathcal{N}_{\lambda}^{(\alpha)}$, respectively,
with the coefficient $A_\lambda$ and ${\hat{\vect{q}}_i=\vect{q}_i/q_i}$.
Equation~\eqref{3NpotOlam} denotes the product of the two tensors forming the scalar
${[\cdots\otimes\cdots]_{00}}$.
The explicit expressions of all quantities appearing in Eqs.~\eqref{3Npotsigq} and~\eqref{3NpotOlam},
including the function $w_{\mathrm{pro}}^{(\alpha)}$,
are presented in~\ref{secIrrTenDec3NF}, 
along with the standard expressions of the three terms of the chiral 3NF at N$^2$LO.
Equation~\eqref{3Npotsigq} is based on the decomposition of the three-body potentials rather than the corresponding matrix elements. In contrast, the spin-tensor decomposition~\cite{ELLIOTT1968241,KIRSON1973110,PhysRevC.45.662} of two-body matrix elements (2BMEs) is widely used for analyzing tensorial structure of 2NFs.

Table~\ref{tabrank} summarizes the tensorial structure of the chiral 3NF at N$^2$LO,
i.e., the allowed values of $\lambda$ are listed for each term of the 3NF.
\begin{table}[!t]
 \begin{center}
  \caption{Allowed rank $\lambda$ of the chiral 3NF at N$^2$LO.
  The contact term consists of only the rank-0 (scalar) component, while the $1\pi$ plus contact term
  has the rank-0 and rank-2 (tensor) structure.
  The $2\pi$ term is classified by the LECs; 
  for the $c_1$ and $c_3$ terms, $\lambda=0$, $1$ (vector) and $2$ are allowed,
  whereas the $c_4$ term involves up to the rank-3 components.}
  \begingroup
  \begin{tabularx}{\columnwidth}{YYYYYY}
   \toprule
     & \multirow{2}{*}{ct} & \multirow{2}{*}{$1\pi+\mathrm{ct}$} & \multicolumn{3}{c}{$2\pi$} \\
   \cmidrule{4-6}
     &                     &                                     & $c_1$ & $c_3$ & $c_4$ \\
   \cmidrule{2-6}
    $\lambda$ & $0$ & $0,2$ & $0$--$2$ & $0$--$2$ & $0$--$3$ \\
   \bottomrule
  \end{tabularx}
  \label{tabrank}
  \endgroup
 \end{center}
\end{table}

\begin{figure*}[!t]
\begin{center}
\includegraphics[width=1.0\textwidth,clip]{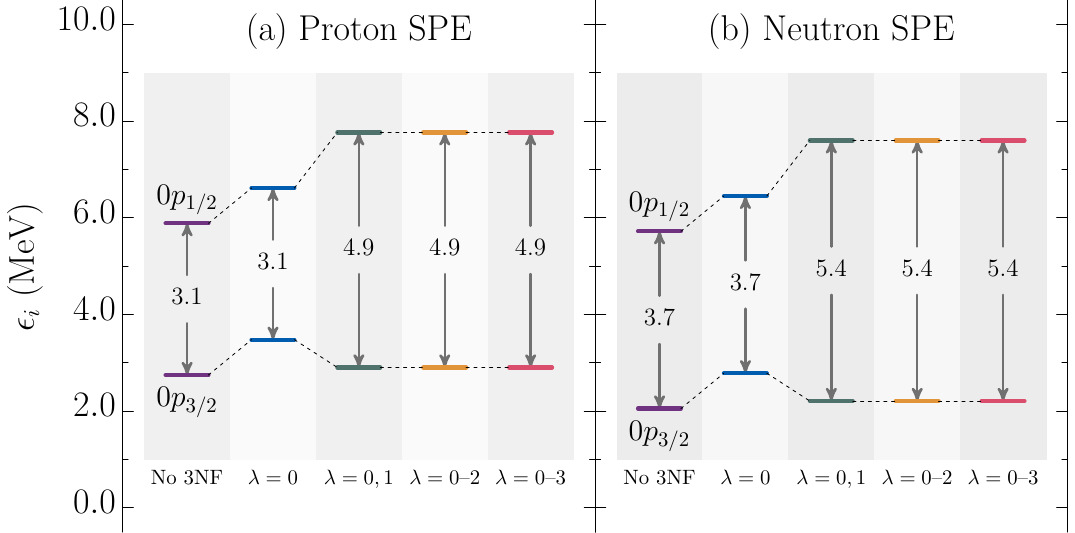}
 \caption{SPEs of the $0p_{3/2}$ and $0p_{1/2}$ states
 for a (a) proton and (b) neutron. The values between the arrows represent the gap of the SPEs.
 The results of ``No 3NF'' are only obtained with the 2NF, 
 whereas the other results include the effect of the 3NF of rank $\lambda$.}
\label{figSPE}
\end{center}
\end{figure*}

The aforementioned decomposition scheme is implemented to investigating the contributions of the different rank components 
of the 3NF on the SO splitting of $p$-shell nuclei within the shell model. 
The role of the whole 3NF in driving the shell evolution in this mass region has been reported in Ref.~\cite{PhysRevC.98.044305}, 
demonstrating its essential contribution to 
improving the description of spectroscopic properties, such as correctly producing
the sequence of the observed states in $^{10}\mathrm{B}$, 
including the ground state ($J^\pi= 3^+$) and the first excited state ($J^\pi= 1^+$) state,
as well as the excitation energy of the yrast state ($J^\pi= 2^+$) in $^{12}\mathrm{C}$.
Herein, $J$ and $\pi$ represent the total spin and parity of nuclei, respectively.

The brief explanation of our framework is as follows:
Our initial Hamiltonian contains the chiral 2NF
at next-to-next-to-next-to-leading order~\cite{PhysRevC.68.041001,MACHLEIDT20111},
along with the Coulomb force for protons,
as well as the chiral 3NF at N$^2$LO.
In accordance with Ref.~\cite{PhysRevLett.99.042501}, we set the values as $c_D=-1.00$ and $c_E=-0.34$, 
which are the LECs associated with the $1\pi$ and contact terms, respectively.
We adopted the nonlocal regulator with the cutoff of $500$~MeV for both 2NF and 3NF.

Moreover, referring to Ref.~\cite{PhysRevC.98.044305}, the effective shell-model Hamiltonian 
for the $0p$-shell valence space outside the doubly magic $^4\mathrm{He}$ was derived within the framework of 
the many-body perturbation theory~\cite{CORAGGIO2009135,CORAGGIO20122125,10.3389/fphy.2020.00345}.
Specifically, the effective shell-model Hamiltonian is expressed 
by way of the Kuo-Lee-Ratcliff folded-diagram expansion~\cite{KUO197165} 
in terms of the $\hat Q$-box vertex function, which was constructed by including
one- and two-body diagrams up to third order for two-body interaction,
as well as one- and two-body diagrams up to first order for the 3NF.
These 3NF diagrams correspond to the normal-ordered one-body (NO1B) and normal-ordered two-body (NO2B) terms 
of the usual expression of the 3NF in the normal-ordered form~\cite{PhysRevLett.109.052501}. 
This so-called normal-ordering approximation 
helps manage the complexity of fully treating the 3NF, 
and has been validated in light- and medium-mass nuclei~\cite{PhysRevLett.109.052501,PhysRevC.90.024312,PhysRevC.93.031301}. 
Within the shell-model framework, the validity of this approximation has been tested 
for $0p$-shell nuclei through benchmark calculations 
comparing shell-model results~\cite{PhysRevC.98.044305} with those from the \textit{ab initio} no-core shell model~\cite{PhysRevC.69.014311,PhysRevLett.99.042501}.
In contrast to Ref.~\cite{PhysRevC.98.044305}, in the present calculations 
we did not include folded diagrams for the 3NF~\cite{CORAGGIO2024104079}, 
i.e., we added the NO1B and NO2B terms of the 3NF, respectively, 
to the single-particle energies (SPEs) and the 2BMEs of the effective shell-model Hamiltonian derived from the 2NF plus Coulomb force.

The omission of the 3NF-folded diagrams implies that changes in the SO splitting induced by the 3NF in systems with one-particle outside a doubly closed core can be directly related to the NO1B term, 
while the NO2B one contributes to the evolution of the SO splitting
as a function of the number of valence nucleons.
In doing so, we neglect some folding contributions related to the 3NF.
However, we have verified that such contributions are not significant 
and do not affect the quantities relevant to the SO splitting, 
which are discussed in this Letter, namely, the SPEs and effective SPEs (ESPEs)~\cite{PhysRevC.74.034330}. 
See~\ref{secESPE} for the definition of the ESPEs.
The little relevance of the 3NF in the folding expansion, 
expressed as products of $\hat Q$-box and its derivatives~\cite{CORAGGIO20122125}, 
is essentially due to the zero value of the first-order-diagram derivatives, 
which do not depend in fact on the energy.

\section{The finding: Importance of rank-1 3NF}
\label{secFind}
First, we discuss the contributions arising from the different rank components of the NO1B term of the 3NF, 
which corresponds to the interactions among one-valence and two-core nucleons 
and affects the SPEs, $\epsilon_i$, of the effective shell-model Hamiltonian. 
Figure~\ref{figSPE} compares the values of $\epsilon_{0p_{3/2}}$ and $\epsilon_{0p_{1/2}}$ derived from the 2NF-only 
with those obtained by gradually adding rank-0 ($\lambda=0$), rank-1 ($\lambda=1$), 
rank-2 ($\lambda=2$), and rank-3 ($\lambda=3$) components of the 3NF.
The 3NF with $\lambda=0$--$3$ corresponds to the whole N$^{2}$LO 3NF. 
Figure~\ref{figSPE} illustrates that the rank-0 component enhances 
$\epsilon_{0p_{3/2}}$ and $\epsilon_{0p_{1/2}}$, 
indicating the repulsive nature of this component,
while leaving the gap between them 
intact.
A significant change in the gap is instead rendered by the rank-1 component, 
which arises exclusively from the $2\pi$-exchange potential (see Fig.~\ref{fig3NFrank}).
This component supplies an attractive and repulsive contribution to
the $0p_{3/2}$ and $0p_{1/2}$ energies, respectively,
significantly enlarging the SO splitting. 
The rank-2 and rank-3 contributions to the SPEs vanish.
The required three-body matrix elements have the form ${\Braket{ 0s_{1/2}0s_{1/2}0p_j | \hat V_{3N} | 0s_{1/2}0s_{1/2}0p_j }}$, where $\hat V_{3N}$ represents the 3NF and $j=1/2,3/2$.
This configuration has $L=L'=1$, where $L$ and $L'$ denote the total orbital angular momenta of the three-nucleon bra and ket states, respectively.
The rank-3 contribution therefore vanishes already at the three-body level, since only $\lambda \leq L+L'=2$ is allowed.
The rank-2 matrix elements can be nonzero, but their contribution vanishes upon contraction over the spin-saturated $^4\mathrm{He}$ core, because the resulting one-body spin operator admits only $\lambda \leq 1/2+1/2=1$.
Hence, the SO splitting induced by the chiral 3NF is primarily governed by its rank-1 component.
These SPEs are comparable with experimental spectra of $^5\mathrm{He}$ and $^5\mathrm{Li}$.
However, the precise position of the $1/2^-$ state relative to the ground state remains uncertain~\cite{PhysRevC.55.536,TILLEY20023}.

The mechanism through which the rank-1 3NF enlarges the gap of the SPEs is determined by the characteristic of the $c_3$ term of the $2\pi$-exchange process.
The $c_3$ potential with $\lambda=1$ depends on
${\left(\vect{\tau}_i\cdot\vect{\tau}_j\right)\left(\vect{\sigma}_i\times\vect{\sigma}_j\right)\cdot\left(\hat{\vect{q}}_i\times\hat{\vect{q}}_j\right)\left(\hat{\vect{q}}_i\cdot\hat{\vect{q}}_j\right)}$,
which is given by Eq.~\eqref{3NpotOlam_c3_vector},
and generates the one-body SO potential, 
as expressed by Eq.~(4.23) of Ref.~\cite{10.1143/PTP.66.227}.
In particular, this explanation also applies to the SO splitting observed in the ESPEs reported herein.

\begin{figure}[!t]
\begin{center}
\includegraphics[width=0.48\textwidth,clip]{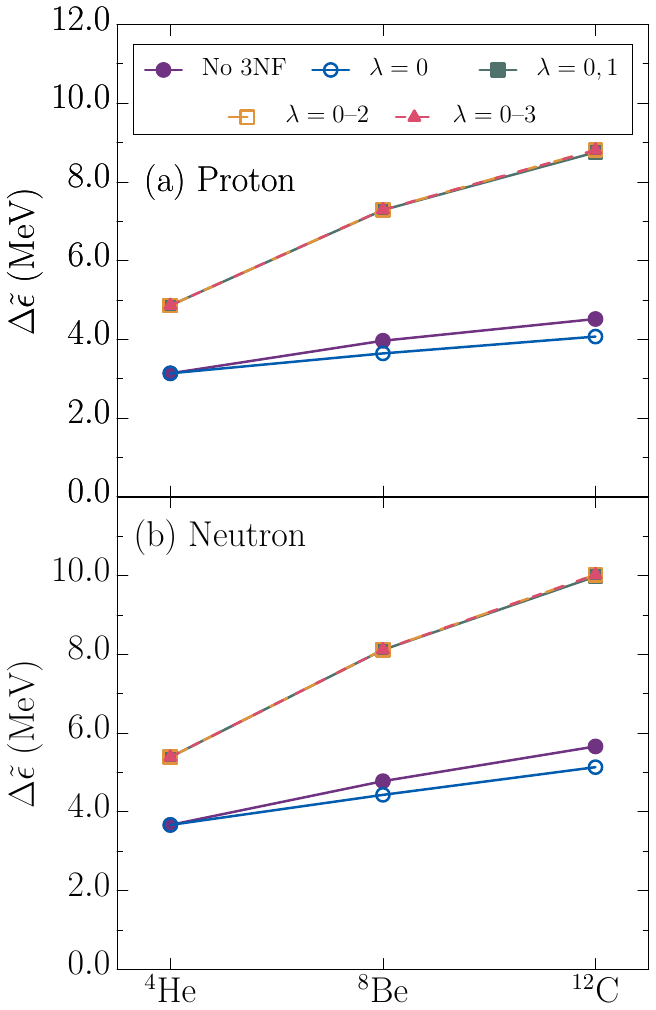}
 \caption{Gap of the ESPEs between $0p_{3/2}$ and $0p_{1/2}$ states for a (a) proton and (b) neutron
 as a function of the mass number of the self-conjugate nuclei.
 While the purple-filled circles are the results obtained with the 2NF, the blue-open circles, green-filled squares, 
 yellow-open squares, and red-filled triangles
 are those with 2NF plus 3NF of the rank $\lambda$ up to 0, 1, 2, and 3, respectively.}
\label{figESPE}
\end{center}
\end{figure}
Subsequently, using the SPEs reported above, 
we consider the evolution of the $0p$-SO splitting.
With this perspective, we investigate the ESPEs,
which represent the single-particle states 
generated by the mean field incorporating the effects from other nucleons outer the inert core.
In fact, changes in the ESPEs induced by the effects of the 3NF 
on the monopole component of the interaction are crucial to describe the observed 
spectroscopic properties of nuclei (see Ref.~\cite{CORAGGIO2024104079} and references therein).
The occupation numbers required in the calculation of the ESPEs are reported in Table~\ref{tabocc} in~\ref{secESPE}; 
these values are obtained in the shell-model framework using the {\sc kshell} code~\cite{SHIMIZU2019372}.

Figure~\ref{figESPE} depicts the SO splitting $\Delta\tilde\epsilon$, defined as the ESPE gap 
between the $0p_{3/2}$ and $0p_{1/2}$ orbitals,
for even--even self-conjugate $0p$-shell nuclei.
As indicated by the difference between the purple-filled circles and red-filled triangles,
the energy gap is enhanced by the whole 3NF, and the enhancement intensifies as the system becomes heavier.
In $^{12}\mathrm{C}$, the 3NF exhibits the most pronounced effect, 
approximately doubling the original $\Delta\tilde\epsilon$ computed with the 2NF only.
This doubling is consistent with that observed in $^{15}\mathrm{N}$~\cite{PhysRevLett.70.2541},
for which the Urbana-VII 3NF~\cite{SCHIAVILLA1986219} was implemented to the variational cluster Monte Carlo method.
Although similar trends are observed in $^{4}\mathrm{He}$ and $^{8}\mathrm{Be}$, the magnitude of the effect diminishes.

We investigated the contribution of the different rank components of the 3NF.
Similarly to the aforementioned SPEs, Fig.~\ref{figESPE} reveals that 
the rank-0 3NF (blue-open circles) plays a 
minor role in changing the SO splitting.
The primal impact comes from the rank-1 3NF (green-filled squares) of the $2\pi$-exchange process, 
accounting for almost 
100\% of the $^8\mathrm{Be}$- and $^{12}\mathrm{C}$-gap enhancements by the whole 3NF.
The 
effects 
of the rank-2 3NF (yellow-open square)
rank-3 3NF 
are negligible.
%
%
%
%
%
%
%
%
%
%
%

Although the rank-2 components enter both 1$\pi$ and $2\pi$ terms,
our numerical analysis revealed that the dominant rank-2 contributions to $\Delta\tilde\epsilon$ originated from the latter.
Therefore, we corroborate even in the SO splitting the $2\pi$-exchange dominance of the 3NF.
This dominance was also affirmed as the leading term of the Fujita--Miyazawa 3NF~\cite{10.1143/PTP.17.360} 
in the $\Delta$-full chiral EFT~\cite{EPELBAUM200865,MACHLEIDT20111} 
and by numerical calculations of ground-state energies for $sd$-shell nuclei~\cite{MA2020135257,Tsunoda2020}.
We emphasize that our finding will not depend on the choice of the contact LECs for the 2NF and 3NF
since the 2$\pi$ LECs are relatively well constrained by the Roy-Steiner equation analysis of pion-nucleon scattering~\cite{PhysRevLett.115.192301}.

Additionally, for the SO splitting, the $c_1$ term plays a minor role compared to the $c_3$ term, 
even though their potentials are similar with each other as given by Eqs.~\eqref{3NpotOlam_c1_vector} and~\eqref{3NpotOlam_c3_vector}.
This is because the $c_1$ and $c_3$ terms are respectively responsible for the $s$- and $p$-wave pion-nucleon scattering 
in the $2\pi$-exchange 3NF, and the $p$-wave scattering is dominant 
as evidenced by the disappearance of the $s$-wave part in the Fujita--Miyazawa 3NF~\cite{10.1143/PTP.17.360}.
Moreover, we numerically confirmed that the rank-1 component of the $c_3$ term accounts for 
nearly the entire rank-1 contribution, 
indicating the negligible influence of the $c_1$ and $c_4$ terms.
Detailed findings will be reported in an upcoming publication, as they lie currently beyond the scope of this Letter.

\section{Conclusion and perspectives}
\label{secConcl}
We introduced the decomposition of the chiral 3NF at N$^2$LO in terms of the rank of irreducible tensors. Based on this technique, we demonstrated the predominant influence of the rank-1 3NF on the SO splitting in the $p$-shell nuclei. 
This conclusion will not be affected by the choice of the contact LECs
since the rank-1 component originates exclusively from the $2\pi$-exchange 3NF.
In this connection, it would be valuable to investigate whether the rank-1 3NF we use 
can account for specific observables of proton--deuteron scattering,
namely, linear combinations of the vector analyzing powers,
which are particularly sensitive to the rank-1 3NF
as reported in Refs.~\cite{PhysRevC.66.044005,doi:10.1142/9789813230897_0010}.

This study offers a fresh perspective on the mechanisms shaping shell structure driven by nuclear forces.
However, to assess the robustness of our conclusions,
it is essential to extend our analysis to heavier mass regions. 
In a near future, we plan to investigate the contribution 
of the different rank components of the 3NF in $pf$-shell nuclei, 
for which the global effect of the 3NF was already evidenced~\cite{PhysRevC.100.034324}.


In conclusion, we remark on the antisymmetry of the rank-1 3NF.
The antisymmetric nature manifests in the two-body subsystems of the complete three-body system, 
owing to the operator $\vect{\sigma}_i\times\vect{\sigma}_j$ 
of the $c_1$ and $c_3$ terms [see Eqs.~\eqref{3NpotOlam_c1_vector} and~\eqref{3NpotOlam_c3_vector}].
This operator induces singlet--triplet mixing of the spin states in the two-body subsystems, akin to the spin canting of magnetic ions driven by the Dzyaloshinsky--Moriya interaction~\cite{DZYALOSHINSKY1958241,PhysRevLett.4.228,PhysRev.120.91}.
Consequently, the rank-1 3NF can change the quantum entanglement manifested as 
the superposition of the product states of the spin-up and spin-down components.
While quantum entanglement in nuclei has been intensively studied through nucleon--nucleon correlations~\cite{PhysRevD.14.2543,PhysRevLett.97.150405,Kwasniewicz_2014,Kruppa_2021,Johnson_2023,PhysRevC.107.044005,PhysRevC.107.L061602,PhysRevC.108.L031002,Perez-Obiol2023},
exploring the change in quantum entanglement prompted by the 3NF presents an intriguing avenue for future research.

\section*{Acknowledgements}
The authors are deeply grateful to L. Coraggio, N. Itaco, 
Y. Kanada-En'yo, M. Kimura, K. Nomura, Y. Taniguchi, Y. Tanimura, and Y. Yamaguchi for fruitful discussions and useful comments.
This work was supported by JSPS KAKENHI Grant Numbers JP21K13919 and JP23KK0250, 
as well as JST ERATO Grant Number JPMJER2304.
The calculations were carried out using the computer facilities at Yukawa Institute for Theoretical Physics, Kyoto University

\appendix
\section{Irreducible-tensor decomposition of chiral three-nucleon force}
\label{secIrrTenDec3NF}
\subsection{Standard expression of potential}
Here we explicitly show the irreducible-tensor decomposition of the potential of the three-nucleon force (3NF) 
derived within the chiral effective field theory at next-to-next-to-leading order (N$^2$LO).

First, it is useful to give the standard expression of the N$^2$LO-three-nucleon potential $v_{3N}^{(\alpha)}$.
We use the letter $\alpha$ as a representative of each term of potential,
namely, the contact (ct) term, the one-pion ($1\pi$) exchange plus contact term,
or the $c_n$ terms of the two-pion ($2\pi$) exchange process.
Here $c_n$ denotes one of the low-energy constants (LECs), $c_1$, $c_3$, or $c_4$.
The contact potential $v_{3N}^{\rm (ct)}$, the $1\pi$ potential $v_{3N}^{(1\pi)}$, 
and the $2\pi$ potential ${v_{3N}^{(2\pi)}=v_{3N}^{(c_1)}+v_{3N}^{(c_3)}+v_{3N}^{(c_4)}}$,
correspond to the right, middle, and left diagrams of Fig.~\ref{fig3NFrank}, respectively.
Following Refs.~\cite{PhysRevC.49.2932,PhysRevC.66.064001,MACHLEIDT20111}, they are usually expressed by
\begin{align}
 v_{3N}^{\rm (ct)}
 &=\frac{c_E}{2f_\pi^4\Lambda_\chi}\sum_{i \ne j \ne k}
 \vect{\tau}_i\cdot\vect{\tau}_j,
 \label{3NFct}\\
 v_{3N}^{(1\pi)}
 &=
 -\frac{g_Ac_D}{8f_\pi^4\Lambda_\chi}
 \sum_{i \ne j \ne k}  \vect{\tau}_i\cdot\vect{\tau}_j
 \frac{\left(\vect{\sigma}_j\cdot\vect{q}_j\right)\left(\vect{\sigma}_i\cdot\vect{q}_j\right)}{q_j^2+m_\pi^2},
 \label{3NF1pi}\\
 v_{3N}^{(c_1)}
 &=
 -\frac{g_A^2c_1m_\pi^2}{2f_\pi^4}
 \sum_{i \ne j \ne k}  \vect{\tau}_i\cdot\vect{\tau}_j
 \frac{\left(\vect{\sigma}_i\cdot\vect{q}_i\right)\left(\vect{\sigma}_j\cdot\vect{q}_j\right)}
 {\left(q_i^2+m_\pi^2\right)\left(q_j^2+m_\pi^2\right)},
 \label{3NFc1}\\
 v_{3N}^{(c_3)}
 &=
 \frac{g_A^2c_3}{4f_\pi^4}
 \sum_{i \ne j \ne k}  \vect{\tau}_i\cdot\vect{\tau}_j
 \frac{\left(\vect{\sigma}_i\cdot\vect{q}_i\right)\left(\vect{\sigma}_j\cdot\vect{q}_j\right)}
 {\left(q_i^2+m_\pi^2\right)\left(q_j^2+m_\pi^2\right)}
 \vect{q}_i\cdot\vect{q}_j,
 \label{3NFc3}\\
 v_{3N}^{(c_4)}
 &=
 \frac{g_A^2c_4}{8f_\pi^4}
 \sum_{i \ne j \ne k} \left(\vect{\tau}_i\times\vect{\tau}_j\right)\cdot\vect{\tau}_k
 \nonumber\\
 &\times
 \frac{\left(\vect{\sigma}_i\cdot\vect{q}_i\right)\left(\vect{\sigma}_j\cdot\vect{q}_j\right)}
 {\left(q_i^2+m_\pi^2\right)\left(q_j^2+m_\pi^2\right)}
 \left(\vect{q}_i\times\vect{q}_j\right)\cdot\vect{\sigma}_k,
 \label{3NFc4}
\end{align}
where the nucleon spin (isospin) operator is expressed by the Pauli matrices $\vect{\sigma}_i$ ($\vect{\tau}_i$).
The momentum transfer is defined by ${\vect{q}_i=\vect{p}_i'-\vect{p}}$, 
with $\vect{p}_i$ and $\vect{p}_i'$ the initial and final momenta of the nucleon $i$, respectively.
The LECs $c_E$ and $c_D$ are associated with the contact and $1\pi$ terms, respectively.
The above potentials contain also the pion-decay constant $f_\pi$, the chiral symmetry breaking scale $\Lambda_\chi$,
the axial vector coupling constant $g_A$, and the average pion mass $m_\pi$.

All three-nucleon potentials given by Eqs.~\eqref{3NFct} to~\eqref{3NFc4}
can be represented in a general form, as shown by Eq.~\eqref{3Npotgeneral}.
We recall it here:
\begin{align}
 v_{3N}^{(\alpha)}
 &= \sum_{i \ne j \ne k} v^{(\alpha)}\!\left(\vect{\tau}_i,\vect{\tau}_j,\vect{\tau}_k\right)
 w^{(\alpha)}\!\left(\vect{\sigma}_i,\vect{\sigma}_j,\vect{\sigma}_k,\vect{q}_i,\vect{q}_j\right).
 \label{3Npotgeneral_recall}
\end{align}
This equation means that the isospin part $v^{(\alpha)}$
can be separated from the spin-momentum part $w^{(\alpha)}$.
By comparing Eq.~\eqref{3Npotgeneral_recall} and the standard expression
of the three-nucleon potentials, Eqs.~\eqref{3NFct} to~\eqref{3NFc4},
we find the explicit form of $v^{(\alpha)}$:
\begin{align}
 v^{(\mathrm{ct})}\!\left(\vect{\tau}_i,\vect{\tau}_j,\vect{\tau}_k\right)
 &=
 v^{(1\pi)}\!\left(\vect{\tau}_i,\vect{\tau}_j,\vect{\tau}_k\right)
 \nonumber\\
 &=
 v^{(c_1)}\!\left(\vect{\tau}_i,\vect{\tau}_j,\vect{\tau}_k\right)
 \nonumber\\
 &=
 v^{(c_3)}\!\left(\vect{\tau}_i,\vect{\tau}_j,\vect{\tau}_k\right)
 \nonumber\\
 &=
 \vect{\tau}_i\cdot\vect{\tau}_j,
 \label{vtautau}\\
 v^{(c_4)}\!\left(\vect{\tau}_i,\vect{\tau}_j,\vect{\tau}_k\right)
 &=
 \left(\vect{\tau}_i\times\vect{\tau}_j\right)\cdot\vect{\tau}_k.
 \label{vtautautau}
\end{align}

\subsection{General expression of irreducible-tensor decomposition}
The irreducible-tensor decomposition of the three-nucleon potential relies on the following relations:
\begin{align}
 \vect{a}\cdot\vect{b}
 &=
 -\sqrt{3}\left[a_1\otimes b_1\right]_{00},
 \label{scalarprod}\\
 \left(\vect{a}\times\vect{b}\right)_\mu 
 &=
 -i\sqrt{2}\left[a_1\otimes b_1\right]_{1\mu}
 \quad(\mu=0,\pm1),
 \label{vectorprod}
\end{align}
where $\vect{a}$ and $\vect{b}$ are arbitrary three-dimensional vectors,
which correspond to irreducible tensors of rank 1, $a_{1m}$ and $b_{1m}$, respectively.
In general, the tensor product of arbitrary irreducible tensors $\mathcal{M}_{lm}$ and $\mathcal{N}_{l'm'}$ is defined by
\begin{align}
 \left[\mathcal{M}_{l}\otimes \mathcal{N}_{l'}\right]_{\lambda\mu}
 =
 \sum_{mm'}\left( l m l' m' | \lambda \mu \right)
 \mathcal{M}_{lm} \mathcal{N}_{l'm'},
\end{align}
with the Clebsch-Gordan coefficient $(\,.\,.\,.\,.\,|\,.\,.\,)$.
Furthermore, assuming $\vect{a}$ depends on neither spin variables nor differential operators,
it can be represented in terms of the spherical harmonics $Y_{1m}$~\cite{doi:10.1142/0270}:
\begin{align}
 a_{1m}
 =
 \sqrt{\frac{4\pi}{3}}a Y_{1m}(\hat{\vect{a}}),
 \label{vecSH}
\end{align}
where $a=\left|\vect{a}\right|$ and ${\hat{\vect{a}}=\vect{a}/a}$.

Applying Eqs.~\eqref{scalarprod},~\eqref{vectorprod}, and~\eqref{vecSH} to $v_{3N}^{(\alpha)}$
given by Eqs.~\eqref{3NFct} to~\eqref{3NFc4}, 
the spin-momentum potential $w^{(\alpha)}$ can be expressed in terms of irreducible tensors of rank $\lambda$.
As already reported by Eqs.~\eqref{3Npotsigq} and~\eqref{3NpotOlam}, 
the resultant expression of $w^{(\alpha)}$ in a general form reads
\begin{align}
 &w^{(\alpha)}\!\left(\vect{\sigma}_i,\vect{\sigma}_j,\vect{\sigma}_k,\vect{q}_i,\vect{q}_j\right)
 \nonumber\\
 &\quad=
 w_{\mathrm{pro}}^{(\alpha)}(q_i,q_j)
 \sum_{\lambda}
 \mathcal{O}_\lambda^{(\alpha)}\!\left(\vect{\sigma}_i,\vect{\sigma}_j,\vect{\sigma}_k,\hat{\vect{q}}_i,\hat{\vect{q}}_j\right),
 \label{3Npotsigq_recall}\\
 &\mathcal{O}_\lambda^{(\alpha)}\!\left(\vect{\sigma}_i,\vect{\sigma}_j,\vect{\sigma}_k,\hat{\vect{q}}_i,\hat{\vect{q}}_j\right)
 \nonumber\\
 &\quad=
 A_\lambda
 \left[\mathcal{M}_{\lambda}^{(\alpha)}\!\left(\vect{\sigma}_i,\vect{\sigma}_j,\vect{\sigma}_k\right)
 \otimes
 \mathcal{N}_{\lambda}^{(\alpha)}\!\left(\hat{\vect{q}}_i,\hat{\vect{q}}_j\right)\right]_{00}.
 \label{3NpotOlam_recall}
\end{align}
The function $w_{\mathrm{pro}}^{(\alpha)}$ includes the pion propagator for the pion-exchange terms,
while it becomes a constant for the contact term.
The LECs are also involved in $w_{\mathrm{pro}}^{(\alpha)}$.
In Eq.~\eqref{3NpotOlam_recall}, the angular momentum coupling coefficients 
dependent on $\lambda$ is denoted by $A_\lambda$, 
and the irreducible tensors $\mathcal{M}_{\lambda\mu}^{(\alpha)}$ and $\mathcal{N}_{\lambda\mu'}^{(\alpha)}$
of rank $\lambda$ are respectively associated with the spin and momentum parts.
Below, the explicit form of these expressions are given.

\subsubsection{Contact term}
Apparently, $v_{3N}^{\rm (ct)}$ given by Eq.~\eqref{3NFct} does not depend on the spin and momenta.
Hence, based on Eqs.~\eqref{3Npotgeneral_recall} and~\eqref{3Npotsigq_recall}, 
the contact term is expressed as the central force, i.e., the rank-0 irreducible tensor only:
\begin{align}
 &w^{(\mathrm{ct})}\!\left(\vect{\sigma}_i,\vect{\sigma}_j,\vect{\sigma}_k,\vect{q}_i,\vect{q}_j\right)
 =
 \frac{c_E}{2f_\pi^4\Lambda_\chi}.
 \label{vct}
\end{align}

Note that the spin-dependent potential can be also selected as the contact term, provided that the antisymmetry is maintained~\cite{PhysRevC.66.064001}.
This characteristic is referred to as the Fierz rearrangement freedom~\cite{PhysRevLett.116.062501,PhysRevC.96.054003,10.3389/fphy.2019.00245}.
We emphasize that, even if the spin-dependent potential is chosen, only the central force contributes to the contact term at N$^2$LO.

\subsection{One-pion exchange plus contact term}
Similarly to the $1\pi$-exchange potential of the two-nucleon force (2NF),
Eq.~\eqref{3NF1pi} can be decomposed into the central and tensor forces 
corresponding to $\lambda=0$ (rank-0 tensor) and  $\lambda=2$ (rank-2 tensor), respectively:
\begin{align}
 &w^{(1\pi)}\!\left(\vect{\sigma}_i,\vect{\sigma}_j,\vect{\sigma}_k,\vect{q}_i,\vect{q}_j\right)
 \nonumber\\
 &\quad=
 -\frac{g_Ac_D}{8f_\pi^4\Lambda_\chi}
 \frac{q_j^2}{q_j^2+m_\pi^2}
 \sum_{\lambda=0,2}
 \mathcal{O}_\lambda^{(1\pi)}\!\left(\vect{\sigma}_i,\vect{\sigma}_j,\vect{\sigma}_k,\hat{\vect{q}}_i,\hat{\vect{q}}_j\right),
 \label{w1pi}\\
 &\mathcal{O}_\lambda^{(1\pi)}\!\left(\vect{\sigma}_i,\vect{\sigma}_j,\vect{\sigma}_k,\hat{\vect{q}}_i,\hat{\vect{q}}_j\right)
 \nonumber\\
 &\quad=
 \sqrt{4\pi}\left(1 0 1 0 | \lambda 0\right)
 \left[\left[\sigma_1(i)\otimes\sigma_1(j)\right]_{\lambda}\otimes Y_\lambda\!\left(\hat{\vect{q}}_j\right)\right]_{00}.
 \label{3NpotOlam_1pi}
\end{align}
Equation~\eqref{3NpotOlam_1pi} can be expressed in terms of the three-dimensional vectors:
\begin{align}
 \mathcal{O}_\lambda^{(1\pi)}\!\left(\vect{\sigma}_i,\vect{\sigma}_j,\vect{\sigma}_k,\hat{\vect{q}}_i,\hat{\vect{q}}_j\right)
 &=
 \begin{dcases}
  \frac{1}{3}\vect{\sigma}_i\cdot\vect{\sigma}_j \quad &(\lambda=0),\\
  \frac{1}{3}\mathcal{S}_{ij}(\hat{\vect{q}}_j) \quad &(\lambda=2),
 \end{dcases}
 \label{3NpotOlam_1pi_vector}
\end{align}
with the standard tensor operator defined by
\begin{align}
 \mathcal{S}_{12}(\hat{\vect{q}})=3\left(\vect{\sigma}_1\cdot\hat{\vect{q}}\right)
 \left(\vect{\sigma}_2\cdot\hat{\vect{q}}\right) -\vect{\sigma}_1\cdot\vect{\sigma}_2.
 \label{tensopeS12}
\end{align}

\subsection{Two-pion exchange $c_1$ term}
The $c_1$ potential $v_{3N}^{(c_1)}$ given by Eq.~\eqref{3NFc1} consists of the rank-0, rank-1, and rank-2 tensors:
\begin{align}
 w^{(c_1)}\!\left(\vect{\sigma}_i,\vect{\sigma}_j,\vect{\sigma}_k,\vect{q}_i,\vect{q}_j\right)
 &=
 -\frac{g_A^2c_1m_\pi^2}{2f_\pi^4}
 \frac{q_i q_j}{\left(q_i^2+m_\pi^2\right)\left(q_j^2+m_\pi^2\right)}
 \nonumber\\
 &\times
 \sum_{\lambda=0}^2
 \mathcal{O}_\lambda^{(c_1)}\!\left(\vect{\sigma}_i,\vect{\sigma}_j,\vect{\sigma}_k,\hat{\vect{q}}_i,\hat{\vect{q}}_j\right),
 \label{wc1}
 \end{align}
 \begin{align}
 &\mathcal{O}_\lambda^{(c_1)}\!\left(\vect{\sigma}_i,\vect{\sigma}_j,\vect{\sigma}_k,\hat{\vect{q}}_i,\hat{\vect{q}}_j\right)
 \nonumber\\
 &\quad=
 \frac{4\pi}{3}\hat{\lambda}
 \left[\left[\sigma_1(i)\otimes\sigma_1(j)\right]_{\lambda}\otimes 
 \left[Y_1\!\left(\hat{\vect{q}}_i\right)\otimes Y_1\!\left(\hat{\vect{q}}_j\right)\right]_{\lambda}\right]_{00}.
 \label{3NpotOlam_c1}
\end{align}
We use the abbreviation $\hat{\lambda}=\sqrt{2\lambda+1}$.
It is useful to represent Eq.~\eqref{3NpotOlam_c1} in terms of vectors:
\begin{align}
 \mathcal{O}_\lambda^{(c_1)}\!\left(\vect{\sigma}_i,\vect{\sigma}_j,\vect{\sigma}_k,\hat{\vect{q}}_i,\hat{\vect{q}}_j\right)
 &=
 \begin{dcases}
  \frac{1}{3}\left(\vect{\sigma}_i\cdot\vect{\sigma}_j\right)\left(\hat{\vect{q}}_i\cdot\hat{\vect{q}}_j\right) \quad &(\lambda=0),\\
  \frac{1}{2}\left(\vect{\sigma}_i\times\vect{\sigma}_j\right)\cdot\left(\hat{\vect{q}}_i\times\hat{\vect{q}}_j\right) \quad &(\lambda=1),\\
  \frac{1}{3}\mathcal{T}_{ij}\left(\hat{\vect{q}}_i,\hat{\vect{q}}_j\right) \quad &(\lambda=2).
 \end{dcases}
 \label{3NpotOlam_c1_vector}
\end{align}
The generalized tensor operator~\cite{10.1143/PTP.97.587} is defined by
\begin{align}
 \mathcal{T}_{12}(\hat{\vect{q}},\hat{\vect{q}}')
 &=
 \frac{3}{2}[(\vect{\sigma}_1\cdot\hat{\vect{q}})
 (\vect{\sigma}_2\cdot\hat{\vect{q}}')
 +(\vect{\sigma}_2\cdot\hat{\vect{q}})
 (\vect{\sigma}_1\cdot\hat{\vect{q}}')]
 \nonumber\\
 &-(\vect{\sigma}_1\cdot\vect{\sigma}_2)
 (\hat{\vect{q}}\cdot\hat{\vect{q}}').
 \label{genetensopeT12}
\end{align}

\subsection{Two-pion exchange $c_3$ term}
The $c_3$ potential $v_{3N}^{(c_3)}$ given by Eq.~\eqref{3NFc3} differs from $v_{3N}^{(c_1)}$
only in the prefactor and the presence of $\vect{q}_i\cdot\vect{q}_j$ 
responsible for the $p$-wave pion-nucleon scattering.
Therefore, the structure of $w^{(c_3)}$ is similar to that of $w^{(c_1)}$:
\begin{align}
 w_{3N}^{(c_3)}\!\left(\vect{\sigma}_i,\vect{\sigma}_j,\vect{\sigma}_k,\vect{q}_i,\vect{q}_j\right)
 &=
 \frac{g_A^2 c_3}{4f_\pi^4}
 \frac{q_i^2 q_j^2}{\left(q_i^2+m_\pi^2\right)\left(q_j^2+m_\pi^2\right)}
 \nonumber\\
 &\times
 \sum_{\lambda=0}^2
 \mathcal{O}_\lambda^{(c_3)}\!\left(\vect{\sigma}_i,\vect{\sigma}_j,\vect{\sigma}_k,\hat{\vect{q}}_i,\hat{\vect{q}}_j\right),
 \label{wc3}
\end{align}
\begin{align}
 &\mathcal{O}_\lambda^{(c_3)}\!\left(\vect{\sigma}_i,\vect{\sigma}_j,\vect{\sigma}_k,\hat{\vect{q}}_i,\hat{\vect{q}}_j\right)
 \nonumber\\
 &\quad=
 4\pi(-)^{\lambda+1}\hat{\lambda}
 \sum_{l_i l_j}
 \left(1 0 1 0 | l_i 0\right)
 \left(1 0 1 0 | l_j 0\right)
 \begin{Bmatrix}
  l_i & l_j & \lambda \\
  1   & 1   & 1
 \end{Bmatrix}
 \nonumber\\
 &\quad\times
 \left[\left[\sigma_1(i)\otimes\sigma_1(j)\right]_{\lambda}\otimes 
 \left[Y_{l_i}\!\left(\hat{\vect{q}}_i\right)\otimes Y_{l_j}\!\left(\hat{\vect{q}}_j\right)\right]_{\lambda}\right]_{00}.
 \label{3NpotOlam_c3}
\end{align}
We can find the similarity between $\mathcal{O}_\lambda^{(c_1)}$ and $\mathcal{O}_\lambda^{(c_3)}$
by rewriting Eq.~\eqref{3NpotOlam_c3} as
\begin{align}
 &\mathcal{O}_\lambda^{(c_3)}\!\left(\vect{\sigma}_i,\vect{\sigma}_j,\vect{\sigma}_k,\hat{\vect{q}}_i,\hat{\vect{q}}_j\right)
 \nonumber\\
 &\quad=
 \begin{dcases}
  \frac{1}{3}\left(\vect{\sigma}_i\cdot\vect{\sigma}_j\right)\left(\hat{\vect{q}}_i\cdot\hat{\vect{q}}_j\right)^2  &(\lambda=0),\\
  \frac{1}{2}\left(\vect{\sigma}_i\times\vect{\sigma}_j\right)\cdot\left(\hat{\vect{q}}_i\times\hat{\vect{q}}_j\right)\left(\hat{\vect{q}}_i\cdot\hat{\vect{q}}_j\right) &(\lambda=1),\\
  \frac{1}{3}\left(\hat{\vect{q}}_i\cdot\hat{\vect{q}}_j\right)\mathcal{T}_{ij}\left(\hat{\vect{q}}_i,\hat{\vect{q}}_j\right)  &(\lambda=2).\\
 \end{dcases}
 \label{3NpotOlam_c3_vector}
\end{align}

\subsection{Two-pion exchange $c_4$ term}
The $c_4$ potential $v_{3N}^{(c_4)}$ given by Eq.~\eqref{3NFc4} involves up to the rank-3 tensors:
\begin{align}
 w_{3N}^{(c_4)}\!\left(\vect{\sigma}_i,\vect{\sigma}_j,\vect{\sigma}_k,\vect{q}_i,\vect{q}_j\right)
 &=
 \frac{g_A^2 c_4}{8f_\pi^4}
 \frac{q_i^2 q_j^2}{\left(q_i^2+m_\pi^2\right)\left(q_j^2+m_\pi^2\right)}
 \nonumber\\
 &\times
 \sum_{\lambda=0}^3
 \mathcal{O}_\lambda^{(c_4)}\!\left(\vect{\sigma}_i,\vect{\sigma}_j,\vect{\sigma}_k,\hat{\vect{q}}_i,\hat{\vect{q}}_j\right),
 \label{wc4}
\end{align}
\begin{align}
 &\mathcal{O}_\lambda^{(c_4)}\!\left(\vect{\sigma}_i,\vect{\sigma}_j,\vect{\sigma}_k,\hat{\vect{q}}_i,\hat{\vect{q}}_j\right)
 \nonumber\\
 &\quad=
 4\pi i\sqrt{6}\hat{\lambda}
 \sum_{l_i l_j S}
 \hat{S}
 \left(1 0 1 0 | l_i 0\right)
 \left(1 0 1 0 | l_j 0\right)
 \begin{Bmatrix}
  l_i & l_j & \lambda \\
  1   & 1   & S \\
  1   & 1   & 1
 \end{Bmatrix}
 \nonumber\\
 &\quad\times
 \left[\left[\left[\sigma_1(i)\otimes\sigma_1(j)\right]_{S}\otimes \sigma_1(k)\right]_{\lambda}\otimes
  \left[Y_{l_i}\!\left(\hat{\vect{q}}_i\right)\otimes Y_{l_j}\!\left(\hat{\vect{q}}_j\right)\right]_{\lambda}\right]_{00}.
\end{align}
The rank-0 and rank-1 components, $\mathcal{O}_0^{(c_4)}$ and $\mathcal{O}_1^{(c_4)}$, respectively,
can be rewritten as
\begin{align}
 &\mathcal{O}_0^{(c_4)}\!\left(\vect{\sigma}_i,\vect{\sigma}_j,\vect{\sigma}_k,\hat{\vect{q}}_i,\hat{\vect{q}}_j\right)
 \nonumber\\
 &\quad=
 \frac{1}{6}\left[\left(\vect{\sigma}_{i}\times\vect{\sigma}_{j}\right)\cdot\vect{\sigma}_{k}\right]
 \left[1-\left(\hat{\vect{q}}_i\cdot\hat{\vect{q}}_j\right)^2\right],
 \label{2pic4rank0}\\
 &\mathcal{O}_1^{(c_4)}\!\left(\vect{\sigma}_i,\vect{\sigma}_j,\vect{\sigma}_k,\hat{\vect{q}}_i,\hat{\vect{q}}_j\right)
 \nonumber\\
 &\quad=
 \frac{1}{10}\left[
 4\left(\vect{\sigma}_i\cdot\vect{\sigma}_j\right)\vect{\sigma}_k
 -\left(\vect{\sigma}_j\cdot\vect{\sigma}_k\right)\vect{\sigma}_i
 -\left(\vect{\sigma}_k\cdot\vect{\sigma}_i\right)\vect{\sigma}_j
 \right]
 \cdot\left(\hat{\vect{q}}_i\times\hat{\vect{q}}_j\right)
 \nonumber\\
 &\quad\times
 \left(\hat{\vect{q}}_i\cdot\hat{\vect{q}}_j\right).
 \label{2pic4rank1}
\end{align}
Although the other two components can be expressed in terms of the vectors,
we omit them here since they are not the main focus of this Letter
and the resultant forms are cumbersome.

\section{Effective single-particle energy}
\label{secESPE}
The effective single-particle energy (ESPE) of the single-particle state $i$ is 
denoted by $\tilde\epsilon_i$ and defined by~\cite{PhysRevC.74.034330}
\begin{align}
 \tilde\epsilon_i&=\epsilon_i +\sum_{j} v_{ij}^{(\mathrm{mon})} \Braket{\hat{N}_j},
 \label{ESPE}\\
 v_{ij}^{(\mathrm{mon})}
 &=
 \frac{\sum_J (2J+1)\Braket{ij;J|\hat{V}|ij;J}}{\sum_J (2J+1)},
 \label{monoint}
\end{align}
where $\epsilon_i$ and $\Braket{ij;J|\hat{V}|ij;J}$ are, respectively, the single-particle energies 
and the two-body matrix elements of our effective shell-model Hamiltonian, 
which is defined with respect to the core nucleus, 
$^4\mathrm{He}$ in the present calculations. 
As mentioned in the main text, both $\epsilon_i$ and $\Braket{ij;J|\hat{V}|ij;J}$ are microscopically derived, 
including one-body and two-body diagrams up to third order in the 2NF and normal-ordered components of the 3NF. 
As expressed by Eq.~\eqref{monoint}, the monopole interaction $v_{ij}^{(\mathrm{mon})}$, 
which is obtained by averaging on all the allowed values of the total angular momentum $J$, 
represents the monopole contribution of the two-body effective interaction $\hat{V}$.
The quantity $\Braket{\hat{N}_{j}}$ in Eq.~\eqref{ESPE} denotes the ground-state occupation number 
of the single-particle state $j$ for the nucleus under investigation. 
The values employed in this Letter are obtained by diagonalizing the effective shell-model Hamiltonian 
and listed in Table~\ref{tabocc}.
\begin{table}[!t]
  \caption{Occupation numbers $\Braket{\hat{N}_{i}}$ for $^{8}\mathrm{Be}$ and $^{12}\mathrm{C}$.
  The columns denoted by ``No 3NF'' are the results obtained with the 2NF only,
  while the other columns are those including the 3NFs of rank $\lambda$.}
  \begingroup
  \begin{tabularx}{\columnwidth}{LWYYYY}
   \toprule
     & & \multicolumn{2}{c}{Proton} & \multicolumn{2}{c}{Neutron}\\
     & & $0p_{3/2}$ & $0p_{1/2}$ & $0p_{3/2}$ & $0p_{1/2}$\\
   \cmidrule{1-6}
   \multirow{5}{*}{$^8\mathrm{Be}$} & No 3NF & 1.447 & 0.553 & 1.491 & 0.509 \\
                                    & $\lambda=0$      & 1.441 & 0.559 & 1.481 & 0.519 \\
                                    & $\lambda=0,1$    & 1.577 & 0.423 & 1.604 & 0.396 \\
                                    & $\lambda=0$--$2$ & 1.569 & 0.431 & 1.597 & 0.403 \\
                                    & $\lambda=0$--$3$ & 1.569 & 0.431 & 1.597 & 0.403 \\
   \cmidrule{1-6}
   \multirow{5}{*}{$^{12}\mathrm{C}$} & No 3NF & 2.952 & 1.048 & 3.049 & 0.951 \\
                                    & $\lambda=0$      & 2.894 & 1.106 & 2.982 & 1.018 \\
                                    & $\lambda=0,1$    & 3.370 & 0.630 & 3.420 & 0.580 \\
                                    & $\lambda=0$--$2$ & 3.331 & 0.669 & 3.387 & 0.613 \\
                                    & $\lambda=0$--$3$ & 3.331 & 0.669 & 3.387 & 0.613 \\
   \bottomrule
 \end{tabularx}
  \label{tabocc}
  \endgroup
\end{table}

\section*{}
During the preparation of this work the authors used ChatGPT 
in order to improve language and readability. 
After using this tool, the authors reviewed and edited the content as needed 
and take full responsibility for the content of the publication.

\bibliographystyle{elsarticle-num} 
\bibliography{SO3NF}






\end{document}